\documentclass{article}
\usepackage[utf8]{inputenc}

\usepackage{authblk} % for affiliations

\usepackage{tablefootnote}

\usepackage{longtable}

\usepackage{float}
\usepackage{tabularx}
\usepackage{graphicx}
\usepackage{amsmath}

\usepackage{array}
\usepackage{booktabs}
\usepackage{adjustbox}

\title{A Decision Support System for Stock Selection and Asset Allocation Based on Fundamental Data Analysis}

\author[1]{Ali Abrishami \thanks{Corresponding author: abrishami@ce.sharif.edu}}
\author[1]{Jafar Habibi}
\author[2]{AmirAli Jarrahi}
\author[1]{Dariush Amiri}
\author[1]{Mohammadamin Fazli}

\affil[1]{Department of Computer Engineering, Sharif University of Technology, Tehran, Iran}
\affil[2]{Tehran Institute for Advanced Studies, University of Khatam}

\providecommand{\keywords}[1]{\textbf{\textit{Keywords:}} #1}

\begin{document}

\maketitle

\clearpage

\section{Abstract}

Financial markets are integral to a country's economic success, yet their complex nature raises challenging issues for predicting their behaviors.
There is a growing demand for an integrated system that explores the vast and diverse data in financial reports with powerful machine-learning models to analyze financial markets and suggest appropriate investment strategies.
This research provides an end-to-end decision support system (DSS) that pervasively covers the stages of gathering, cleaning, and modeling the stock's financial and fundamental data alongside the country's macroeconomic conditions.
Analyzing and modeling the fundamental data of securities is a noteworthy method that, despite its greater power, has been used by fewer researchers due to its more complex and challenging issues.
By precisely analyzing securities' fundamental data, the proposed system assists investors in predicting stock future prices and allocating assets in major financial markets: stock, bond, and commodity.
The most notable contributions and innovations of this research are:
(1) Developing a robust predictive model for mid- to long-term stock returns, tailored for investors rather than traders,
(2) The proposed DSS considers a diverse set of features relating to the economic conditions of the company, including fundamental data, stock trading characteristics, and macro-economic attributes to enhance predictive accuracy,
(3) Evaluating the DSS performance on the Tehran Stock Exchange that has specific characteristics of small to medium-sized economies with high inflation rates and showing the superiority to novel researches, and
(4) Empowering the DSS to generate different asset allocation strategies in various economic situations by simulating expert investor decision-making.

\keywords{Stock Prediction, Fundamental Analysis, Decision Support Systems, Asset Allocation, Financial Markets, Investment Strategies}

\clearpage

\section{Introduction}

Stock and bond markets are two of the most important financial markets. They play a crucial role in funding businesses to expand their production activities and provide services.
The studies show the essential role of financial markets in economic growth \cite{levine1998stock,fink2003bond}.
Simultaneously, they enable individuals and investors to save and increase the value of their money by participating in economic activities \cite{CHO2010626}. Importantly, these markets protect the value of money against inflation in the long run, ensuring a secure financial future \cite{campbell2009understanding,graham2005intelligent,ONGENA2018257}. Keeping commodities like gold and silver in the portfolio also has advantages for investors. They are effective hedges against inflation and are good options for diversification and lowering portfolio risk \cite{malkiel1999random}.

As we know, not all businesses perform the same economically, and some have advantages in making profits. Moreover, in different economic situations, investing in any financial market has risks and opportunities for investors \cite{malkiel1999random}. Therefore, the existence of a decision support system for evaluating markets' conditions and providing suggestions to investors can be very beneficial.
Every investor should obtain a strategy called asset allocation. Asset allocation is the process of determining the percentage of capital invested in specific financial markets, such as stocks, bonds, commodities, and money.
Asset allocation and diversification help reduce the risk of capital loss by spreading investments across different assets and market sectors while also aiming to optimize returns based on risk tolerance. Three main aspects of diversification are asset classes, market, and investing time \cite{malkiel2009elements}.

Most stock market analysts use technical data analysis as the prediction method \cite{NAZARETH2023119640,LIN2024100864}. Technical analysis uses candlestick and volume data to analyze and forecast the future price of stock \cite{kumbure2022machine}. On the other hand, fundamental analysis studies and investigates businesses' financial reports and then uses these data to predict the stock's future performance. Fundamental analysis has more challenges in extracting, analyzing, and modeling data than technical analysis; hence, very little research uses this method \cite{nti2020systematic}. Nevertheless, the results gained from fundamental approaches are superior to technical ones. We will compare the details of technical and fundamental techniques in section \ref{literature_review}.

In this research, we used financial ratio data and macroeconomic features to build a model for stock return prediction and selecting high-potential stocks. This problem formulation is similar to analyses done by an expert investor in making the decision to buy stocks. To achieve this goal, we needed a robust decision support system that solves the challenges of working with unstructured fundamental data. Also, preparing a diverse and thorough set of features for an accurate presentation of the stock market and economic conditions is a mandatory task that we overcame in this research.
After building the stock prediction model, we proposed an asset allocation model as a decision-support system for helping investors allocate budgets in different financial markets. We included the bond and commodity markets because of their lower risk and fluctuations than the stock market and good sources of diversification.
This work has many applications in stock market prediction and asset allocation for individual and institutional investors.

The rest of this paper is organized as follows: In section \ref{literature_review}, we review the related works about stock market prediction approaches and financial decision support systems. In section \ref{section:Proposed-system-architecture}, we describe the proposed decision support system architecture for end-to-end processes of data collection, data cleaning, and model training of the problem. In section \ref{section:stock-prediction-method} we demonstrate the used feature list with their importance for solving the problem. Also this section formulates the problem and presents the proposed model for predicting stock return, and compares it to selected novel baselines. In section \ref{section:asset-allocation-strategy}, we show the applications of the model for predicting the future trend of the stock market and propose some strategies for asset allocation in financial markets. Finally, we conclude the paper in Section \ref{section:conclusion} and describe future works.

\section{Literature review} \label{literature_review}

\subsection{Stock market prediction}

Financial markets are marketplaces where investors can participate in economic activities. One of the most important financial markets is the stock market, that many disciplines are created, and many researches are conducted to study and explore its underlying behaviors \cite{LIN2024100864}. Some of its seminal researches are Fama’s “Efficient Capital Markets” \cite{fama1970efficient}, Graham’s “Security Analysis” \cite{graham1934security}, Ball’s “An empirical evaluation of accounting income numbers” \cite{ball2013empirical}, and many others.

In recent years, with huge advancements in machine learning and deep learning, much attention has been attracted to predicting the stock market future with advanced techniques \cite{zhang2024deep,sezer2020financial}. Some research papers predicted the stock market index \cite{BHANDARI2022100320,song2021forecasting} and many others predicted the stock's future price \cite{JING2021115019,BALLINGS20157046}.
In this manner, different solutions are proposed that use a broad range of features for prediction. The study of Kumbure et al. \cite{kumbure2022machine} showed that, in general, the number of distinct features is about 2173, and it falls into three main categories: technical indicators, macroeconomic variables, and fundamental indicators.
These studies cover a broad range of countries with specific economic behaviors and sizes, spanning different countries in Asia \cite{tsai2010combining,phuoc2024applying,nourbakhsh2023combining}, North America \cite{CHUDZIAK2023119203,YANG2024114213,ZOU2024122801}, South America \cite{DEOLIVEIRACAROSIA2021115470}, and Europe \cite{BALLINGS20157046,stoean2019deep}.

\subsection{Decision support systems in financial markets}

Decision support systems (DSS) play an essential role in helping individuals make decisions by collecting, processing, and analyzing data.
Applications of DSSs cover a broad range of fields, including agriculture, education, healthcare, etc \cite{eom2006survey}.
One of the most important applications of DSS is in finance, which became more popular in recent years \cite{DONG2024114190,KWON2024114100,CHIANG2016195}. 

Kou proposed a system based on quantitative and qualitative data for deciding between buy, keep, or sell options. The quantitative data includes past stock trades data, and qualitative data is gathered from expert traders of the stock market \cite{kuo1998decision}.
Solares et al. architectured a DSS comprised of three main parts: Price forecasting, stock selection, and portfolio optimization subsystems. The price forecasting model is based on fundamental analysis, and the formed portfolio is optimized by investor risk \cite{SOLARES2022118485}.
Cho designed a DSS that provides multi-level and interactive capabilities for investors to predict market index and stock returns \cite{CHO2010626}.

\subsection{Technical analysis}

As we mentioned, a line of research is based on technical analysis of stock prices. With this approach, stock price prediction is done using OHLC price data along with the volume of trades \cite{malkiel1999random,kumbure2022machine}. 
Some previous works used only raw data from price time series, but most technical analysis researches also used technical indicators such as SMA, WMA, RSI, MACD, and others
\cite{nti2020systematic,LIN2024100864,9220868}. In this manner, the selected features (price data with calculated technical indicators) are used with a machine learning method for future return prediction. The most used methods are deep Learning, SVM, and Decision Tree \cite{9220868,olorunnimbe2023deep,nti2020systematic}.

Albahli et al. \cite{ALBAHLI2023120903} used 18 technical indicators alongside OHLC and stock volume data for future stock price prediction. The adopted deep learning method is based on Dense Convolutional Network architecture (DenseNet), which is optimized for stock price prediction on short and long-term horizons. Phuoc et al. \cite{phuoc2024applying} studied the Vietnam stock market, an emerging economy with specific characteristics. They utilized three technical indicators (SMA, MACD, and RSI) and price data for predicting stock price trends and showed that the LSTM model has high accuracy for this task.

Li and Bastos \cite{9220868} provided an extensive review of research papers that adopted deep learning and technical data analysis for predicting stock prices. The study showed that the LSTM model is the most used approach and that deep learning is dominant in predicting stock prices.

\subsection{Fundamental analysis} \label{section-fundamental-analysis}

From another point of view, fundamental analysts say that past prices are insufficient for predicting future stock returns, and the underlying business performance should be considered and analyzed to assess stock performance accurately \cite{malkiel1999random,graham1934security}. So, the stock selection process will be done using business financial reports, including balance sheets, income statements, and statements of cash flow reports \cite{nti2020systematic,tsai2010combining}.

Financial ratios play an important role in fundamental analysis. These formulas quantify a stock's business performance, making comparing different stocks feasible.
Arkan et al. \cite{arkan2016importance} proposed a method that used 12 financial ratios to predict stock prices for 15 companies in the Kuwait stock market. The result was that ROA, ROE, and net profit ratio are the most important features for predicting stock price. The three studied sectors were industrial, services, and investment sector types.
Nakayama et al. \cite{10.1007/978-3-319-93034-3_22} proposed an approach that predicts the one-month stock return for the Japanese stock market using deep learning methods. In their method, five previous months' data with 25 features composed of financial ratios are fed to a regression model, and the one-month return is predicted. They also proposed a strategy for trading stocks in the market.
Tsai et al. \cite{tsai2023stock} predicted the stock return for the next quarter using 18 financial ratios. Five financial ratio categories were liquidity, leverage, probability, market value, and asset efficiency ratios. For stock return prediction, the model was a regression model. Other researchers, such as Kheradyar \cite{kheradyar2011stock}, studied the Malaysian stock market and used fundamental analysis and financial ratios to predict future returns.

\subsection{Comparing technical and fundamental analysis}

If we want to compare two established fundamental and technical approaches, the following items are achieved \cite{petrusheva2016comparative,malkiel1999random,LIN2024100864}:

\begin{enumerate}
\item According to Fama's Efficient Market Hypothesis theory, if the market is in weak form, technical analysis data have no information for traders to predict stock price.

\item The number of studies associated with fundamental approaches is much lower than that of technical approaches. Reasons for this are harder data gathering, semi-structured data format, more complex modeling of fundamental data, and so on. This issue creates an opportunity for more work on fundamental data analysis.

\item In the long term, the stock price is determined by business activity performance.

\item Many known best-performer investors consider fundamental analysis as the method for selecting stock.
\end{enumerate}

\section{Decision support system architecture} \label{section:Proposed-system-architecture}

As previously discussed, fundamental analysis approaches have more complexity and challenges than technical analysis approaches. The challenges lie in gathering, cleaning, integrating, and exploring published fundamental data. Solares et al. used a four-stage architecture that includes data preprocessing, return forecasting, stock selection, and portfolio optimization \cite{SOLARES2022118485}. Similar architectures are adopted by other works for data collection, feature engineering, strategy formulation, and performance evaluation \cite{CHIANG2016195,DONG2024114190,BANIK2022107994}. Thus, a robust system architecture is needed to tackle these challenges.

Similar to the well-established stock markets, companies listed in the Tehran Stock Exchange (TSE) are obliged to publish their financial reports regularly in each quarter. The three main financial reports are the balance sheet, income statement, and statement of cash flow. These reports are available for traders in different formats: PDF, Excel, and HTML.
The most challenging problems for accessing data are:

\begin{itemize}
\item A crawler is needed for continuous download of announcements.
\item Data is published in semi-structured data, populated with texts and numbers.
\item Data format is changed in different quarters, thus requiring a data mapper for unifying formats.
\item Stocks in different industries have different report formats.
\end{itemize}

We need a system architecture to mitigate the mentioned challenges and the capability of crawling online and historically published fundamental data.
The proposed system architecture is a pipeline architecture with the following components and steps:

\begin{description}

\item[1) Data Gathering:] The data gathering step includes crawling and saving reports' announcements (online and historical data) in a SQL database. After downloading announcements, the next step is saving three types of financial reports' data in their original format (HTML files).
\item[2) Data Cleaning:] In this step, the main activity is cleaning of the semi-structured reports, including normalizing texts and numbers, mapping reports to a unified format, and etc. For later access to the cleaned data, we save it in JSON format in the SQL database.
\item[3) Data Loading:] The third step is for loading data from stored JSONs and merging sequential reports published quarterly. This stage speeds up the data processing and acts as a caching system.
\item[4) Stock Modeling:] The main part of the system is training a model that can predict stock return in various periods. For this purpose, different categories of features are prepared for predicting stock return and a machine learning model is trained.
\item[5) Asset Allocation:] The last step is for allocating asset in different securities. The trained model in previous step is used for predicting the stock return and then determining the percentage of fund for each asset. 

\end{description}

\begin{figure}[h!]
\begin{center}
  \includegraphics[width=0.8\textwidth, angle=0]{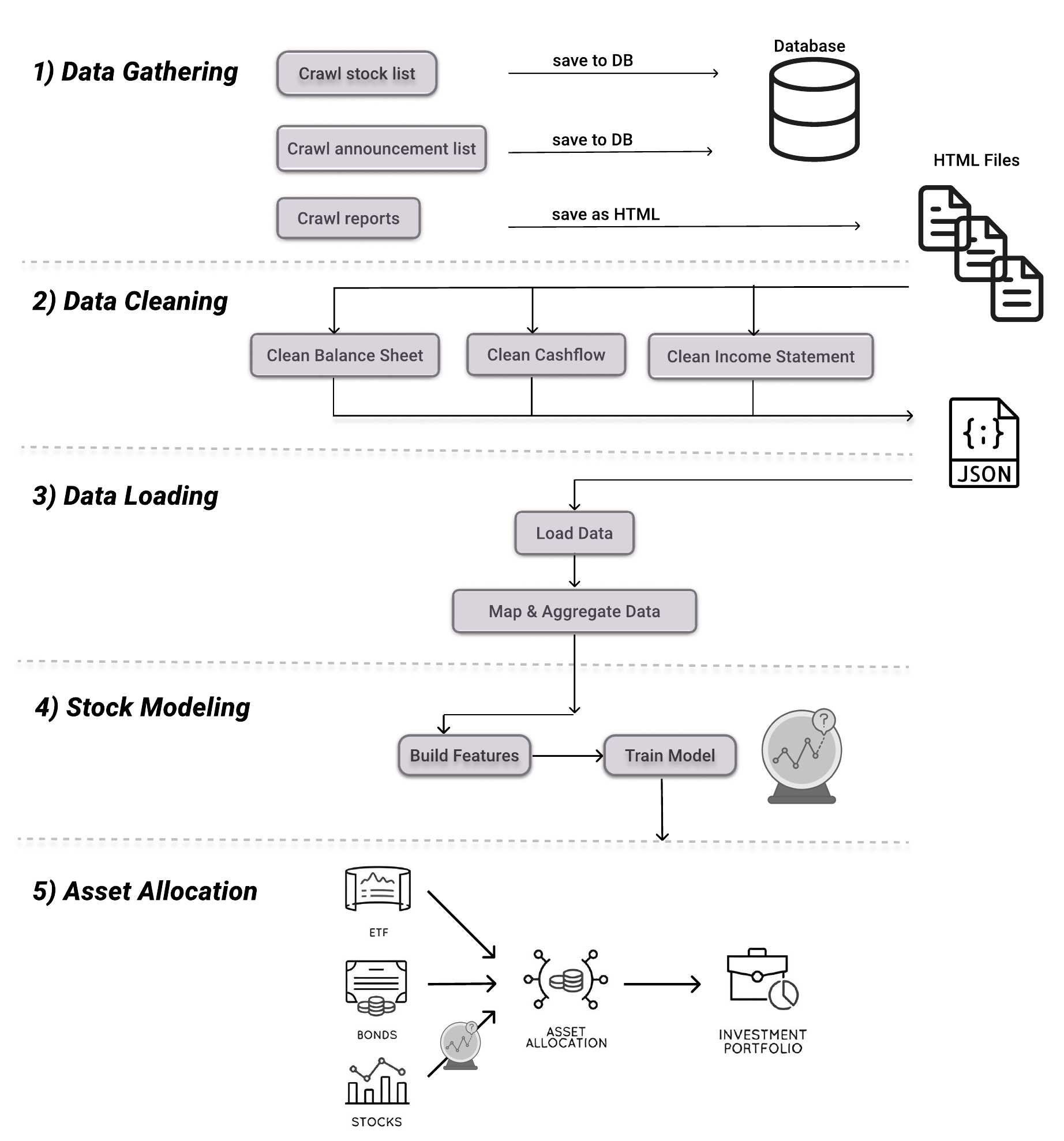}
  \caption{Proposed system architecture}
  \label{fig:system-architecture}
\end{center}
\end{figure}

The main point of system design is the \textbf{pipeline architecture}, which allows the output of each stage to be stored independently, and each stage can be run separately. Stocks have different formats according to their industry, and reports' formats vary over time, so the proposed architecture overcame this problem. Figure \ref{fig:system-architecture} presents the detailed system diagram.

\section{Stock prediction method} \label{section:stock-prediction-method}

In the previous section, we discussed the decision support system architecture. In this section, we will describe the method for predicting future price growth and choosing the best stocks based on their current state. One important step is the process of preparing the feature set. The collected feature set represents a profile of the business and its trading behavior along with the macroeconomic and market factors for a complete representation of the real world. We will describe this in detail in the following subsection.

\subsection{Building feature set}

Feature engineering and feature selection is an essential stage in any machine learning task, specifically stock prediction \cite{10.1007/978-3-642-42042-9_91,htun2023survey}.
We reviewed the most used features and approaches in section \ref{literature_review} and had a comparison between them.
This section presents the methodology and process used to prepare the data and features, along with the rationale for the decision.

As discussed, many experts believe that a company's financial performance, as reported in financial statements, plays a significant role in driving its stock price. Financial ratios are formulas that enable traders to compare stocks to each other and compare stock performance in different periods. Many researchers based their stock prediction on financial ratios that were explained in section \ref{section-fundamental-analysis}.

At first, we build a feature set based on financial ratios for predicting stock return. These ratios are used by expert fundamental analysts to analyze a company's financial performance and health. Number of distinct features is 23, that two most important of them are \textit{Gross Profit Growth} and \textit{Revenue Growth}, which are adopted from \textbf{CANSLIM} method. The features are presented in table \ref{table:financial_ratio_features}.

\renewcommand{\arraystretch}{1.4} % Increase row height for better readability

\begin{longtable}[H]{m{0.5\textwidth} >{\centering\arraybackslash}m{0.45\textwidth}}

\hline
\textbf{Feature} & \textbf{Formula} \\ [0.5ex]
\hline
\hline
\endfirsthead

\text{Debt-to-Equity Ratio} &
$\frac{\text{Total Debt}}{\text{Total Equity}}$
\\[0.5ex]
\hline

\text{Return on Fixed Assets} &
$\frac{\text{Net Income}}{\text{Average Fixed Assets}}$
\\[0.5ex]
\hline

\text{Debt Ratio} & 
$\frac{\text{Total Debt}}{\text{Total Assets}}$
\\[0.5ex]
\hline

\text{Gross Profit Margin} &
$\frac{\text{Gross Profit}}{\text{Revenue}}$
\\[0.5ex]
\hline

\text{Current Ratio} &
$\frac{\text{Current Assets}}{\text{Current Liabilities}}$
\\[0.5ex]
\hline

\text{Net Income Margin} &
$\frac{\text{Net Income}}{\text{Revenue}}$
\\[0.5ex]
\hline

\text{Operating Profit Margin} &
$\frac{\text{Operating Profit}}{\text{Revenue}}$
\\[0.5ex]
\hline

\text{Interest Coverage Ratio} &
$\frac{\text{Earnings Before Interest and Taxes}}{\text{Interest Expense}}$
\\[0.5ex]
\hline

\text{Return-on-Equity Ratio} &
$\frac{\text{Net Income}}{\text{Average Shareholders’ Equity}}$
\\[0.5ex]
\hline

\text{Cash Flow to Income} &
$\frac{\text{Operating Cash Flow}}{\text{Net Income}}$
\\[0.5ex]
\hline

\text{Quick Ratio} &
$\frac{\text{Current Assets - Inventory}}{\text{Current Liabilities}}$
\\[0.5ex]
\hline

\text{Long-Term Debt Ratio} &
$\frac{\text{Long-Term Debt}}{\text{Total Assets}}$
\\[0.5ex]
\hline

\text{Return on Assets} &
$\frac{\text{Net Income}}{\text{Average Total Assets}}$
\\[0.5ex]
\hline

\text{Inventory Turnover} &
$\frac{\text{Cost of Goods Sold}}{\text{Average Inventory}}$
\\[0.5ex]
\hline

\text{Asset Turnover} &
$\frac{\text{Revenue}}{\text{Average Total Assets}}$
\\[0.5ex]
\hline

\text{Cash Ratio} &
$\frac{\text{Cash + Cash Equivalents}}{\text{Current Liabilities}}$
\\[0.5ex]
\hline

\text{Gross Profit Growth} &
$\frac{\text{Current Period Gross Profit}}{\text{Previous Year Same Period Gross Profit}} - 1$
\\[0.5ex]
\hline

\text{Revenue Growth} &
$\frac{\text{Current Period Revenue}}{\text{Previous Year Same Period Revenue}} - 1$
\\[0.5ex]
\hline

\text{RE/TA Ratio} &
$\frac{\text{Retained Earnings}}{\text{Total Assets}}$
\\[0.5ex]
\hline

\text{P/E TTM} &
$\frac{\text{Market Price per Share}}{\text{Earnings per Share}}$
\\[0.5ex]
\hline

\text{P/E TTM Median} &
The median of P/E ratio over the trailing twelve months
\\[0.5ex]
\hline

\text{P/S TTM} &
$\frac{\text{Market Price per Share}}{\text{Revenue per Share}}$
\\[0.5ex]
\hline

\text{P/S TTM Median} &
The median of P/S ratio over the trailing twelve months
\\[0.5ex]
\hline

\text{Beta (1 year)} &
$\frac{\text{Cov(Rm,Rs)}}{\text{Var(Rs)}}$
\\[0.5ex]
\hline

\caption{Financial ratio features}
\label{table:financial_ratio_features}

\end{longtable}

The impact of financial ratios may vary depending on the stock's industry and type. Two main stock type features are the stock industry and whether the stock is a production company or not. By adding these two features, the model performance increases because of considering the effect of financial ratio in relation to stock type.
The stock type features are presented in table \ref{table:Stockـtypeـfeatures}.

\begin{table}[H]
\centering
\begin{tabularx}{\textwidth}{m{0.5\textwidth} m{0.45\textwidth}}

\hline
\textbf{Feature} & \textbf{Description} \\
\hline
\hline

Stock Industry & Stock's industry and sector\\
\hline

Stock Market Exchange & The market exchange for stock\\
\hline

Activity Type & Company activity type including production and others\\
\hline

\end{tabularx}

\caption{Stock type features}
\label{table:Stockـtypeـfeatures}

\end{table}

By using financial ratio and stock type features, the financial performance of the company will be analyzed well. However, for accurate prediction of stock prices, we need to include market data features about stock and market conditions. These features show the investor's tendency to buy stock and invest in it. These features with descriptions are presented in table \ref{table:stock_trading_features}.

\begin{table}[H]
\centering
\begin{tabularx}{\textwidth}{m{0.5\textwidth} m{0.45\textwidth}}

\hline
\textbf{Features} & \textbf{Calculation} \\
\hline
\hline

Average Price Volatility & The average daily volatility of price in the last month. Volatility computed as High/Low \\
\hline

Average Daily Return & The average daily return of stock in the last month\\
\hline

Average Trades Value & The average daily value of stock trades in the last month\\
\hline

B/S Power Ratio & The daily average of the power divided by cellular power in the last month. power is the value individual traded  divided by number of individual traded. \\
\hline

Ownership Change & The average value individual bought and sold in the last month.\\
\hline

\end{tabularx}     

\caption{Stock trading features}
\label{table:stock_trading_features}

\end{table}

The last main category of features is about macroeconomic indicators. The good performance of macroeconomic indicators significantly impacts stock prices. So the features listed in table \ref{table:macro_market_features}, appended to the features list, that had a good effect on model accuracy.
In Iran's capital market, the strength of the national currency against other currencies, specifically the dollar, is essential data in examining the company's performance and the attractiveness of the capital market.

\begin{table}[H]
\centering
\begin{tabularx}{\textwidth}{m{0.50\textwidth} m{0.45\textwidth}}

\hline
\textbf{Feature} & \textbf{Description} \\
\hline
\hline

Gov. Bonds Return & Government Bonds Last Month Average Return \\
\hline

USD/IRR Exchange Rate & The value of the U.S. dollar (USD) against the Iranian rial (IRR) \\
\hline

USD/IRR Exchange Rate Return & The USD/IRR last month return \\
\hline

Equal-Weight Index Return & Equal-Weight Index last month return \\
\hline

Market Index Value & The market capitalization weighted index of TSE \\
\hline

Market Index Return & Weighted index three months return \\
\hline

Gold Spot USD-Return & 1 month return \\
\hline

\end{tabularx}

\caption{Macroeconomic and market features}
\label{table:macro_market_features}

\end{table}

\begin{figure}[h!]
\begin{center}
    \includegraphics[width=0.7\textwidth, angle=0]{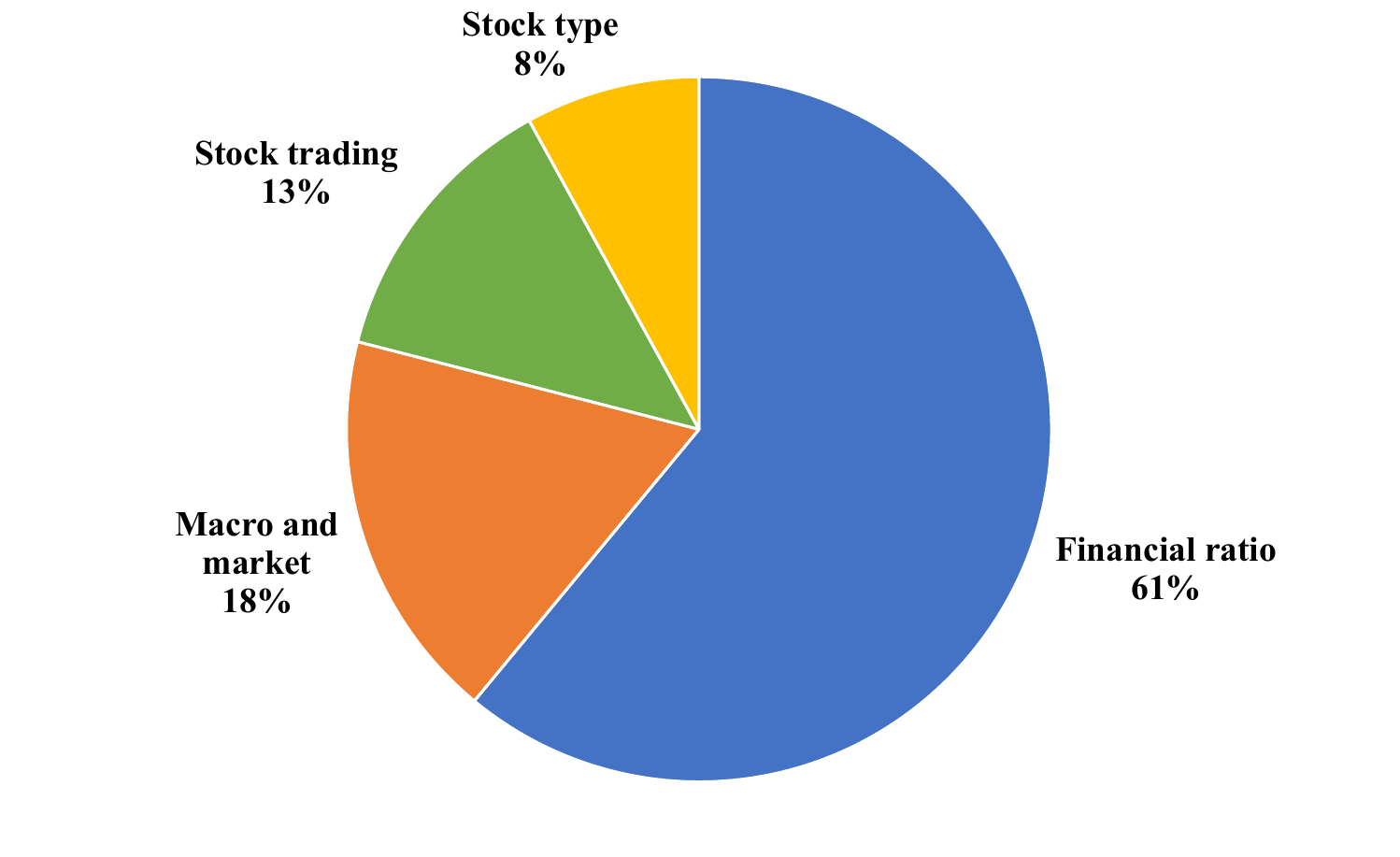}
    \caption{Distribution of features by feature type}
    \label{fig:ratios-distribution}
\end{center}
\end{figure}

So far, we prepared multiple feature categories that indicate the company's financial performance to predict stock return. We tried to model an expert analyst's workmanship in studying stock and market conditions. The distribution of feature types is illustrated in figure \ref{fig:ratios-distribution}.

\subsection{Problem formulation}

The stock return prediction is generally formulated as follows: Given the past information about stock trades and financial condition, the stock's expected return in time period t is estimated.
This problem is solved as a classification (e.g. \cite{BALLINGS20157046, tsai2010combining}) or regression problem (e.g. \cite{nourbakhsh2023combining, RATHER20153234, phuoc2024applying}) for predicting future stock price.

In this research, we formulated the problem in a somewhat different way. The traders can gain a return proportional to the inflation rate by investing in low-risk government bonds. In this regard, the problem formulation is done in a manner that investing in the specified stock could produce more return than fixed-income ETF or not.
If the estimated return is higher than that of a fixed-income ETF, the system recommends buying the stock. Otherwise, it is better to invest in low-risk markets.
Fixed-income ETFs are a type of ETFs which most of their assets are composed of government bonds and securities such as stocks. By investing in this type of asset, the investor can avoid losses and gain profits in the market downturn. The monthly return computation is presented in the formula \ref{eq:monthly-return}.

\begin{equation} \label{eq:monthly-return}
\textit{Monthly return} = (1 + YTM) ^ {(1/12)}
\end{equation}

\subsection{Prediction model}

Stock market traders evaluate stock conditions by analyzing the financial features of stock and predicting the probability of price growth. In the same way, the adopted model is a an artificial neural network model that learns the relationship between computed features and the future return of stock. We designed the model output as a two-class prediction problem that label 1 is for stock with higher return than fixed-income ETF in the same period. This problem modeling has the advantage that the trader can decide to choose between stock market or a less risky asset such as fixed-income with limited known return. Also, this modeling is very beneficial in countries with high inflation rates. If investing in the stock market is not as profitable as the bond market, the investor will prioritize the bond market.

The model output is a probability number between 0 and 1 that higher probability shows the model's confidence about future stock return.
The artificial neural network contains one hidden layer with 100 neurons. The Adam optimizer selected as the optimization method and binary cross entropy for the loss function.
The batch size parameter is set to 32, and the model is trained in 50 epochs.

\begin{equation}
\textit{Binary Cross Entropy} = -\frac{1}{N} \sum_{i=1}^{N} \left( y_i \cdot \log(\hat{y}_i) + (1 - y_i) \cdot \log(1 - \hat{y}_i) \right)
\end{equation}

The feature set is normalized with the standard scaler for better model convergence.

\subsection{Dataset}

\begin{figure}[h!]
\begin{center}
    \includegraphics[width=0.7\textwidth, angle=0]{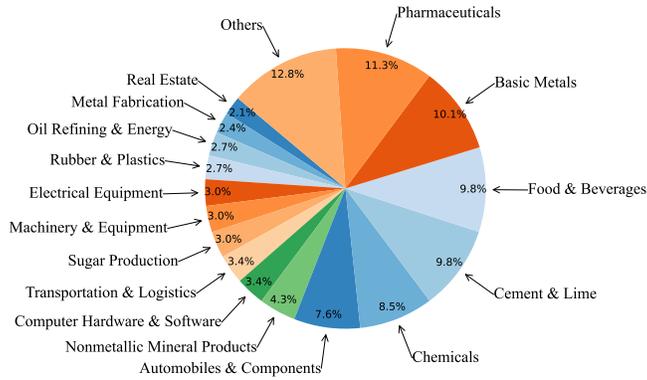}
    \caption{Industry share of total market symbols}
    \label{fig:industry-number-of-stocks}
\end{center}
\end{figure}

\begin{figure}[h!]
\begin{center}
    \includegraphics[width=0.7\textwidth, angle=0]{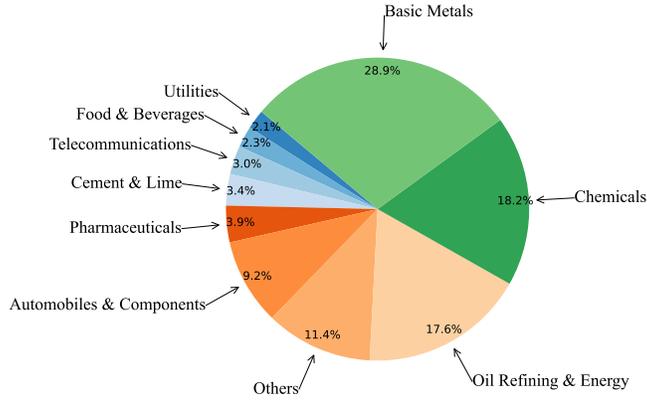}
    \caption{Market capitalization distribution across industry sectors}
    \label{fig:industry-market-cap}
\end{center}
\end{figure}

The dataset consists of 413 stocks in the Tehran Stock Exchange (TSE) that publish financial reports regularly. The industry share of total market symbols and market capitalization distribution across industry sectors are presented in figures \ref{fig:industry-number-of-stocks} and \ref{fig:industry-market-cap} respectively. For stock return prediction, we used adjusted stock prices (that were normalized with regard to capital increase and DPS). The collected data covers the period from 2015 to 2024. For the training set, we used the \%75 of the data sorted by the earliest publish date, and the other \%25 of the data was used for the testing data.

The publishing of financial reports has a one-month lag from the scheduled time, and we incorporated this delay in simulations and experiments. With this regard, we have a more accurate simulation of the real world.

\subsection{Results}

For the evaluation of the proposed method, we selected three similar scientific researches that used fundamental data for stock price prediction.
These researches are Tsai et al. \cite{tsai2023stock}, Abe et al. \cite{10.1007/978-3-319-93034-3_22}, and Arkan \cite{arkan2016importance} works that we name Baseline-1, Baseline-2, and Baseline-3 respectively which discussed with details in section \ref{section-fundamental-analysis} of literature review.
The common point of these studies is the use of financial ratio data, which we compared with our proposed method by using their features and training a model for stock prediction. Also, we added five other machine learning models named Support Vector Machine (SVM), Random Forest (RF), Decision Tree (DT), K-Nearest Neighbours (KNN), and Logistic Regression (LR) as other approaches.

For analyzing model accuracy, we chose 1, 2, 3, 4, 5, 6, 9, and 12-month periods for stock return prediction. Table \ref{table:stock_prediction} presents the accuracy of our proposed method and baselines.

\begin{table}[H]

\begin{tabularx}{\textwidth}{X c c c c c c c c}

    \hline
    Model &
    1 &
    2 &
    3 &
    4 &
    5 &
    6 &
    9 &
    12 \\

    \hline
    \hline
    Proposed  (\%)       & 64 & 65 & 79 & 79 & 73 & 74 & 67 & 61 \\
    \hline
    Baseline-1 (\%)       & 68 & 58 & 50 & 58 & 59 & 58 & 56 & 58 \\
    \hline
    Baseline-2 (\%)       & 70 & 61 & 63 & 59 & 58 & 59 & 63 & 60 \\
    \hline
    Baseline-3 (\%)       & 68 & 58 & 61 & 60 & 60 & 59 & 56 & 58 \\
    \hline
    SVM (\%)     & 70 & 70 & 57 & 63 & 62 & 57 & 55 & 49 \\
    \hline
    RF (\%)      & 65 & 58 & 61 & 62 & 62 & 58 & 53 & 48 \\
    \hline
    DT (\%)      & 51 & 60 & 61 & 61 & 57 & 59 & 47 & 51 \\
    \hline
    KNN (\%)            & 57 & 57 & 55 & 59 & 55 & 56 & 52 & 48 \\
    \hline
    LR (\%)             & 62 & 31 & 41 & 41 & 39 & 43 & 42 & 39 \\
    \hline

\end{tabularx}

\caption{Performance comparison of the proposed method with the baselines (The columns period are in months)}
\label{table:stock_prediction}

\end{table}

The average accuracy of the proposed method and baselines on the train and test datasets are presented in table \ref{table:stock_prediction_average}.
The superior power of the proposed method indicates the better modeling of the stock forecasting problem, which in summary results from the following reasons:

\begin{enumerate}
\item The better modeling of market and macroeconomic features that influence stock price and adopting more powerful and more diverse features that explore the market behavior.
\item Using more data samples in the broader time range enables the model to explore different market trends and various sectors.
\item Modeling the problem as a classification task and forecasting the price in comparison to fixed-income ETF return helps to better investigate stock direction and strategy development.
\end{enumerate}

\begin{table}[H]

\begin{tabularx}{\textwidth}{X c c c c c c c c}

    \hline
    Model &
    Train Accuracy &
    Test Accuracy  \\

    \hline
    \hline
    Proposed  (\%)       & 88 & 70 \\
    \hline
    Baseline-1 (\%)       & 60 & 58 \\
    \hline
    Baseline-2 (\%)       & 77 & 62 \\
    \hline
    Baseline-3 (\%)       & 60 & 50 \\
    \hline

\end{tabularx}

\caption{Average accuracy of proposed method and baselines on the train and test data}
\label{table:stock_prediction_average}

\end{table}

\begin{figure}[h!]
\begin{center}
    \includegraphics[width=0.8\textwidth, angle=0]{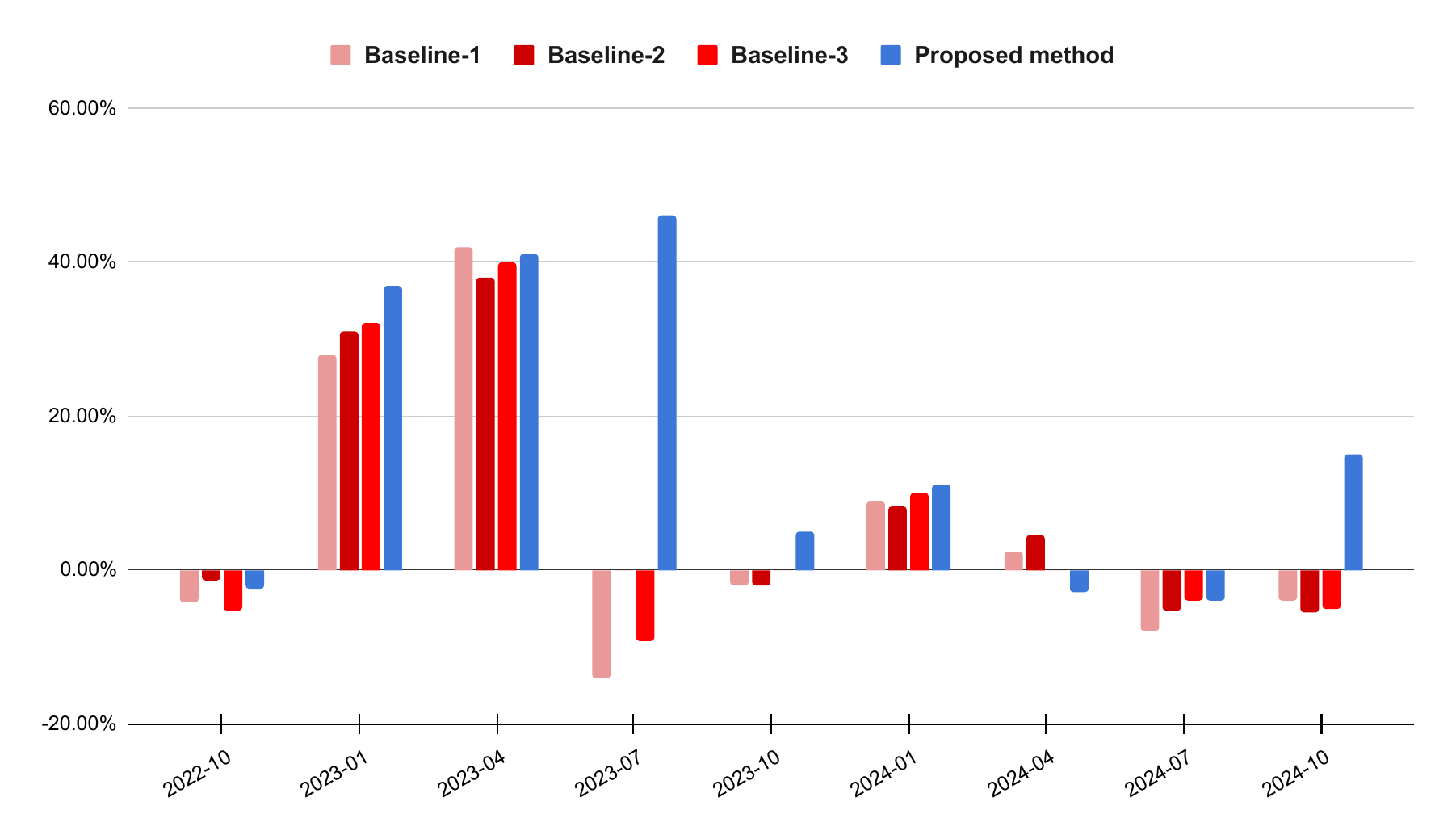}
    \caption{Methods return in three month periods}
    \label{fig:strategy-3-months}
\end{center}
\end{figure}

\begin{figure}[h!]
\begin{center}
    \includegraphics[width=0.8\textwidth, angle=0]{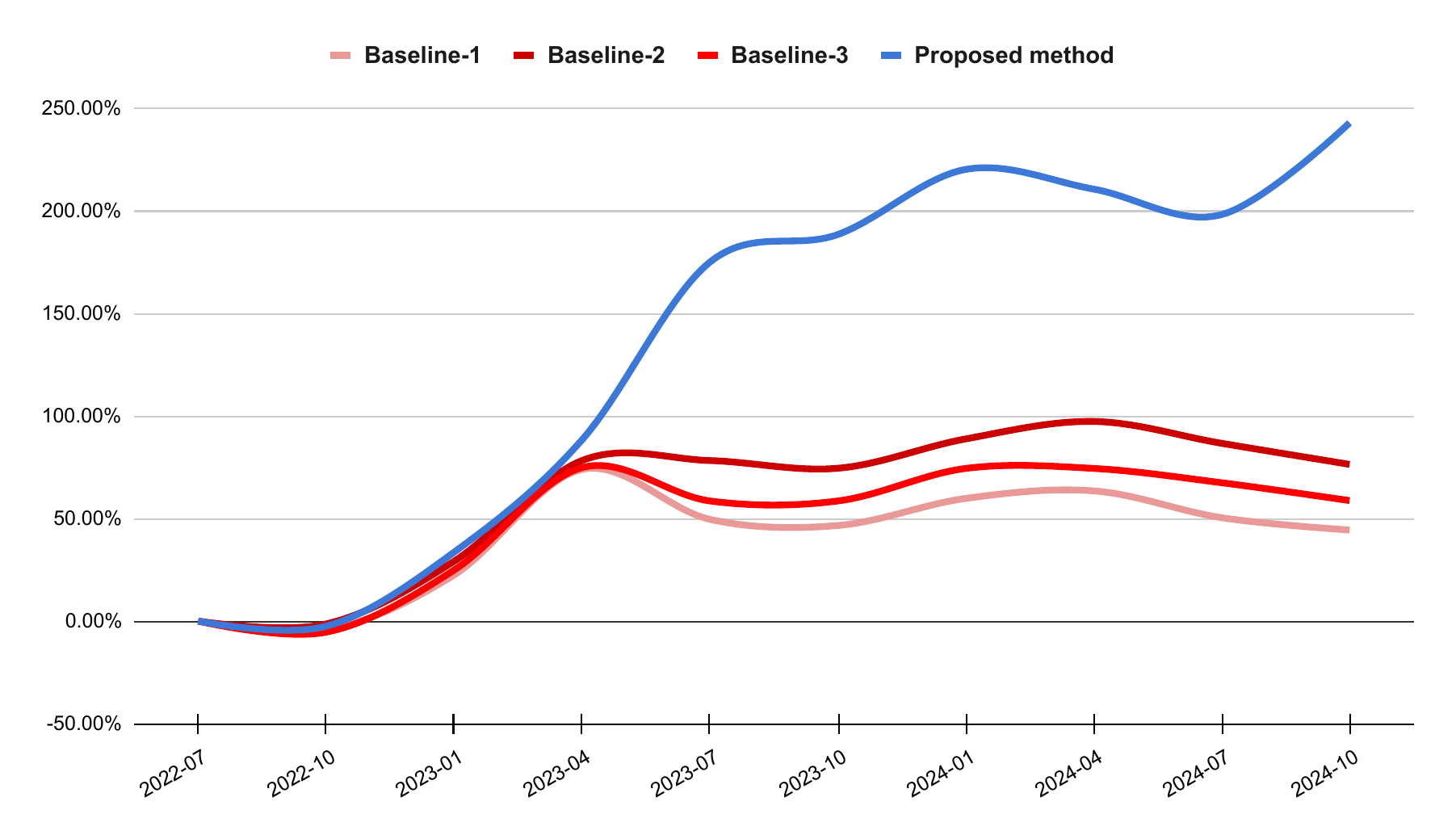}
    \caption{Cumulative return of methods in test period}
    \label{fig:strategy-compounded-return}
\end{center}
\end{figure}

Abe \cite{10.1007/978-3-319-93034-3_22} and Tsai \cite{tsai2023stock} proposed trading strategies by forming a portfolio consists of stocks with the highest predicted return. Tsai's proposed strategy selects 20 high-return stocks, and hence, we designed a similar experiment for comparison between our proposed method and baselines:
In the 3-month periods of publishing financial reports of stocks, select 20 high-return stocks of each method and then form a portfolio with these stocks. 
The results of portfolio periodic return is presented in figure \ref{fig:strategy-3-months} for each method. Also, the cumulative return of each portfolio is presented in figure \ref{fig:strategy-compounded-return}. The results show the superiority of the proposed method to baselines.

\section{Asset allocation strategy} \label{section:asset-allocation-strategy}

Asset allocation plays a central role in financial economics \cite{wachter2010asset} and helps investors decide capital distribution in different assets. A straightforward strategy is to distribute a 60/40 portfolio between stocks and fixed-income instruments \cite{durall2022assetallocationmarkowitzdeep}. In this section, we first propose a method for predicting the stock market direction, and based on the proposed method, we suggest two strategies for building a portfolio.

\subsection{Stock market movement prediction}\label{Stock-market-movement-prediction}

The stock market's health is a key indicator of overall economic well-being, and the good performance of stocks attracts investor's attention to the stock market among different financial markets. Many works tried to predict the stock market index to help the traders for making decision of \cite{BHANDARI2022100320,song2021forecasting,kumbure2022machine}.
Hence, one of the main goals of this study is on predicting the stock market's return.

The stock market consists of stocks of various sizes spanning different economic sectors. Cho \cite{CHO2010626} tried to predict the stock market direction by predicting all stocks in the market. Our proposed approach is a similar approach for predicting market performance and is as follows:

\begin{enumerate}
    \item Predict the \( P(s_{i}) \) for stock \( s_{i} \) that is, the probability of the stock return exceeds the fixed-income return with the prediction model described in section \ref{section:stock-prediction-method}.
    \item Compute \( P(market) \) with all stocks in the market by the formula \ref{eq:market-index-probability}. \( P(market) \) is the probability of the stock market return as a whole exceeds the fixed-income return. The market cap considers the size of the company in market return calculation.
\end{enumerate}

\begin{equation}\label{eq:market-index-probability}
P(market) = \frac{\sum\limits_{i=1}^{N} P(s_{i}) * \textit{market cap}(s_{i})} {\sum\limits_{i=1}^{N} \textit{market cap}(s_{i})}
\end{equation}

As a result, we have an estimate of market growth based on company financial performance and the country's macroeconomic conditions.

\begin{figure}[h!]
\begin{center}
    \includegraphics[width=1.0\textwidth]{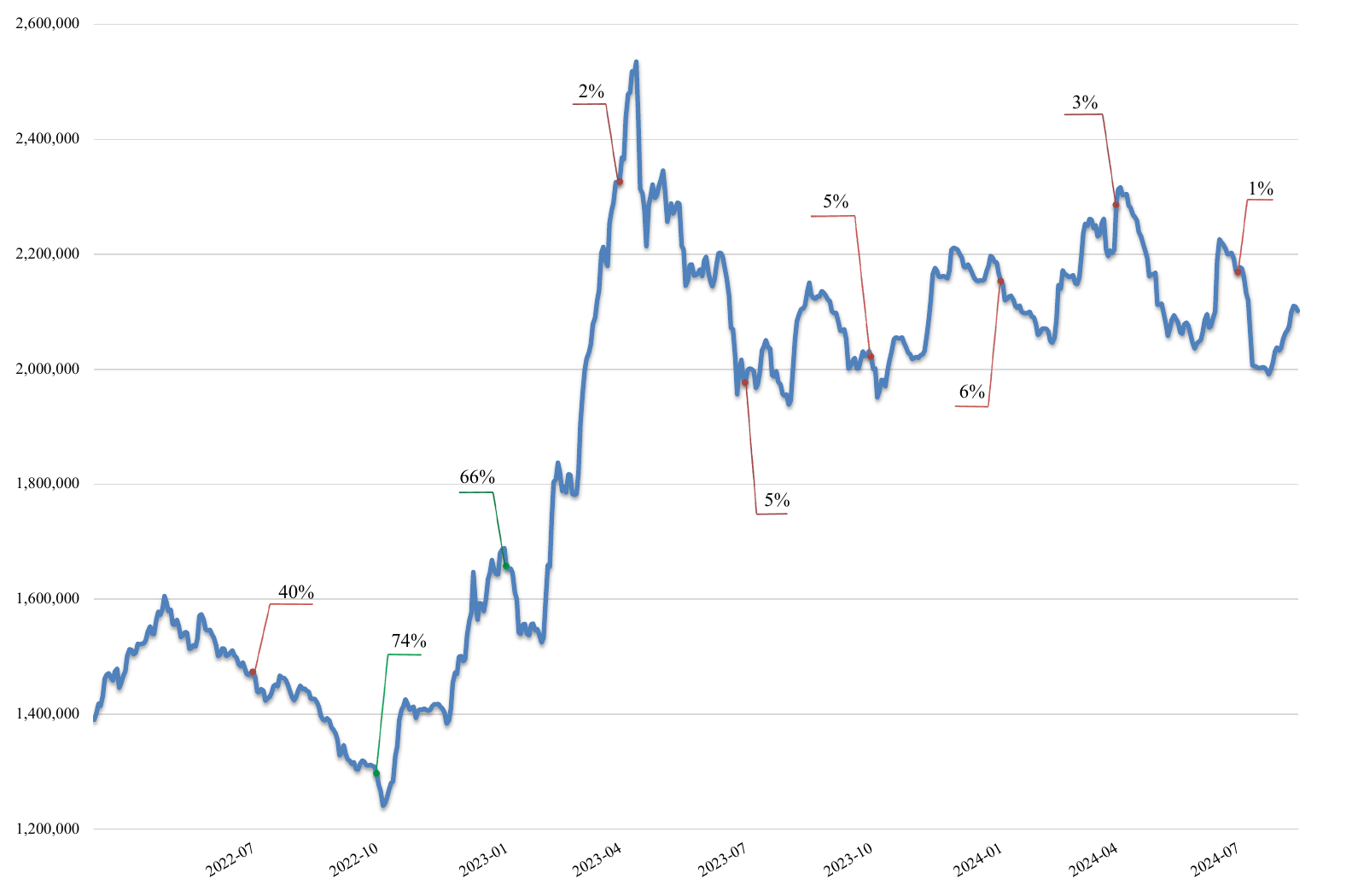}
    \caption{Predicted probabilities for market return exceed fixed-income return in test period}
    \label{fig:result-market-index}
\end{center}
\end{figure}

In figure \ref{fig:result-market-index}, we presented the Tehran stock market index in the test period, and the predicted probabilities for market return exceed fixed-income return. As seen, the model performance is good in different conditions and has good accuracy for predicting future trends, which enables traders to make investment decisions.

In the following, we will present proposed trading strategies based on the predicted expected market growth.

\subsection{Allocating asset in selected financial markets}

After the explanation of market trend predictor model, we will describe the asset allocation component of the decision support system in this section.
Asset allocation, in general, is the decision of distributing funds in different assets of various financial markets. 
As stated in the introduction, asset allocation in various markets is a key decision for every investor. With asset diversification, the investment risk is minimized in terms of the earned return.

The three main assets that individual investors should have in their portfolios are stocks, bonds, and gold. 
Gold is one of the most important commodities that hedges against inflation and maintains its value. Also, because of its low correlation with stock and bond, it is a good choice for portfolio diversification that many principal approaches recommend for a part of investment portfolio \cite{malkiel1999random}. The last point about gold asset is its good performance in geopolitical tensions and financial crises for hedging against economic uncertainty. 
Bonds are another important asset, that are long-term debt instruments issued by governments and corporations. Bonds have lower risk compared to stocks, that will be very beneficial for downward trends of the stock market. We included fixed-income ETFs in the experiments that is composed of different types of bonds and have risk and return similar to bonds.
Stocks are the last asset we included in the portfolio. We selected leveraged stock ETFs for the stock portion of the portfolio.

The system's core functionality is based on the proposed stock market return predictor described in section \ref{Stock-market-movement-prediction}. At first, the probability of growth of stock market return compared to fixed-income ETFs is computed. In this manner, we have a rough probability of stock market growth compared to the bond market and a prediction of the market index.

Based on the computed weighted probabilities, we choose between two asset allocation strategies:

\begin{enumerate}
\item Market growth probability of more than \%50
\item Market growth probability of less than \%50
\end{enumerate}

The percentage of assets allocated to different markets is presented in table \ref{table:strategy_results}.

\begin{table}[H]

\begin{tabularx}{\textwidth}{m{0.2\textwidth} m{0.35\textwidth} >{\centering\arraybackslash}m{0.2\textwidth} >{\centering\arraybackslash}m{0.15\textwidth}}

    \hline
    Asset type &
    Description &
    Scenario 1 &
    Scenario 2 \\
    
    \hline
    \hline
    Gold & Gold ETF & \%20 & \%20 \\ 
    \hline
    Bond & Fixed-income ETF  & \%10  & \%70 \\ 
    \hline
    Stock & Market index ETFS  & \%70 & \%10 \\ 
    \hline

\end{tabularx}

\caption{Assets weight in two scenarios}\label{table:strategy_results}

\end{table}

\begin{figure}[h!]
\begin{center}
  \includegraphics[width=1\textwidth, angle=0]{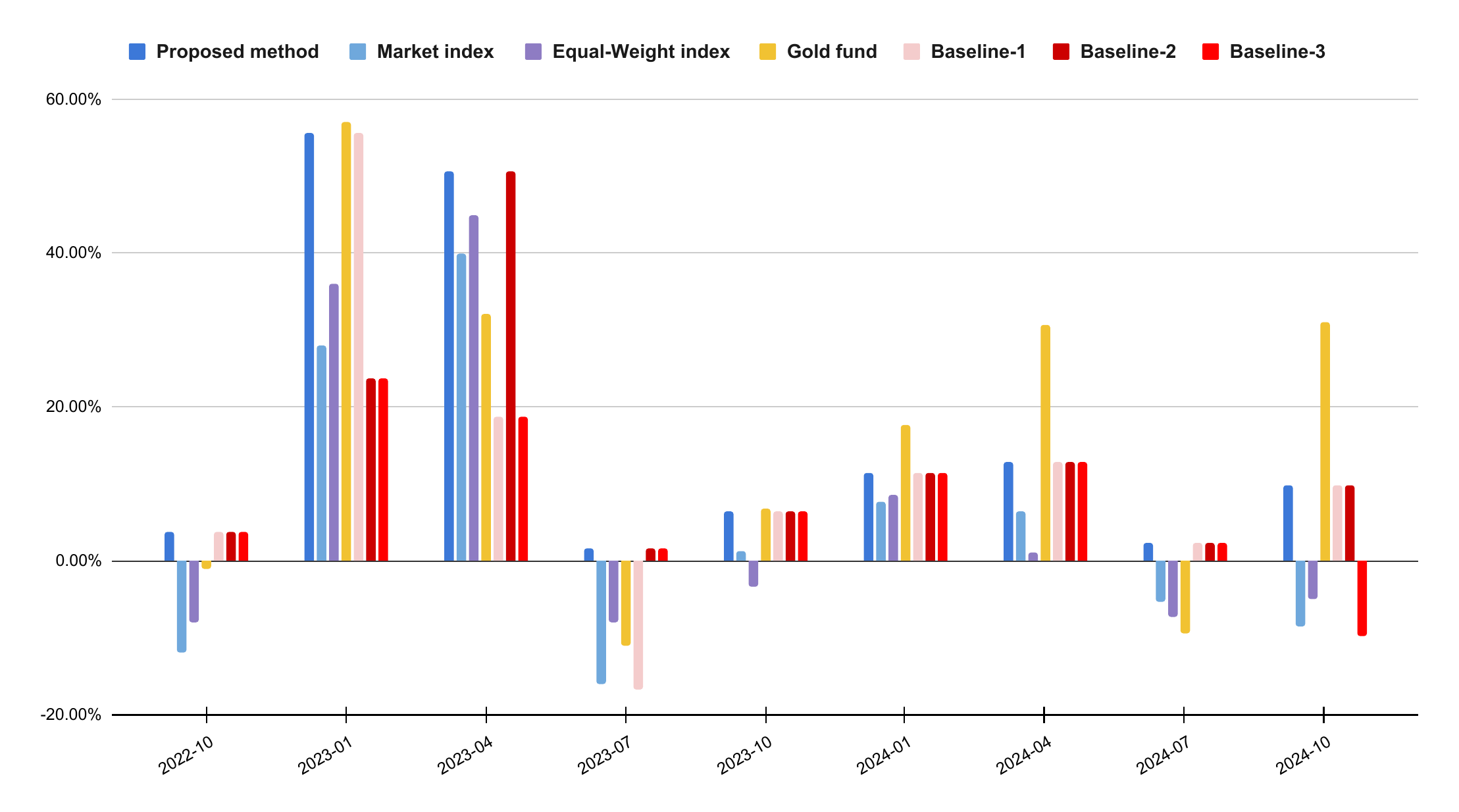}
  \caption{Three months nominal return of strategies in test period}
  \label{fig:results-3-months-return}
\end{center}
\end{figure}

\begin{figure}[h!]
\begin{center}
  \includegraphics[width=1\textwidth, angle=0]{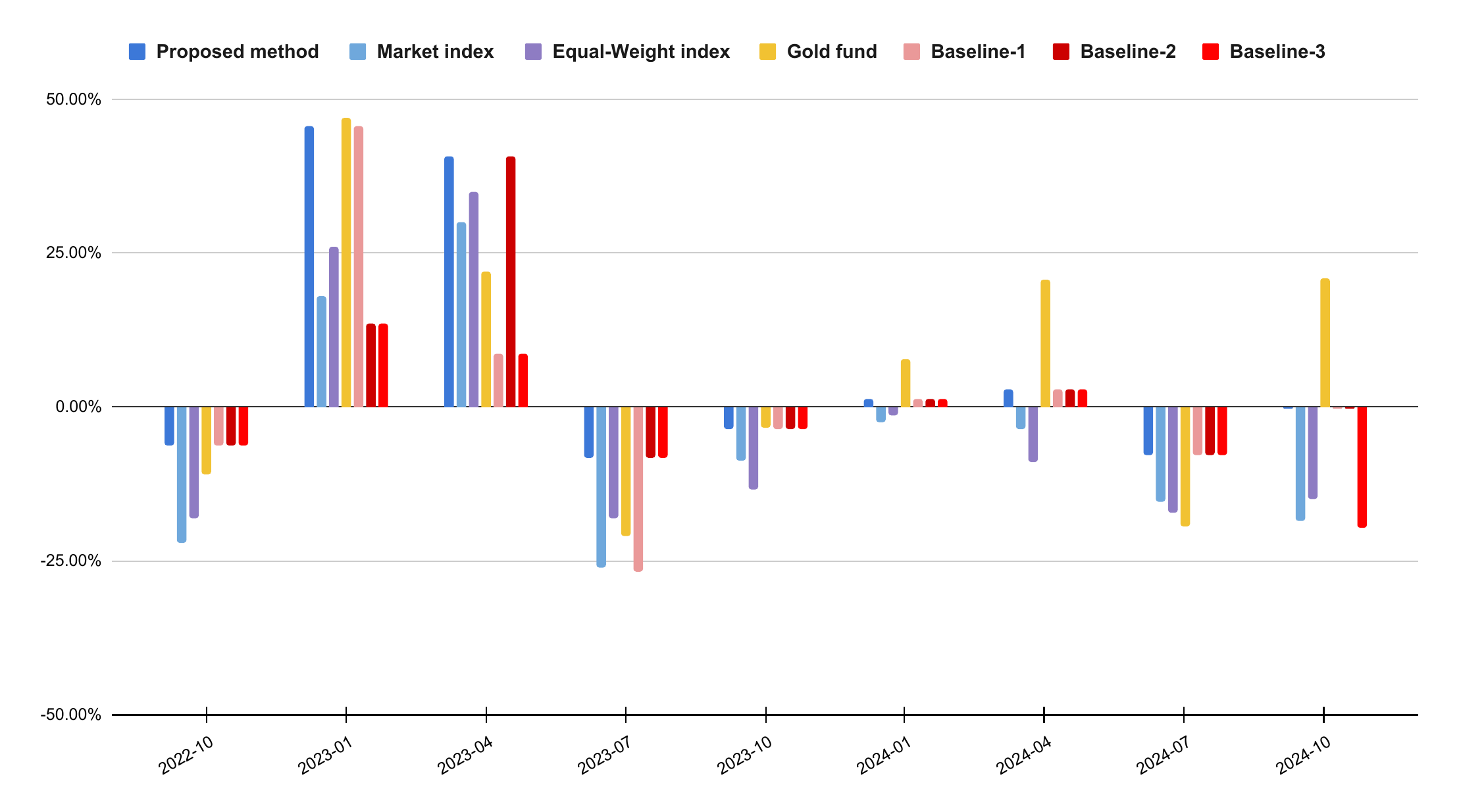}
  \caption{Three months real return of strategies in test period}
  \label{fig:resutls-3-months-return-adjusted}
\end{center}
\end{figure}

\begin{figure}[h!]
\begin{center}
  \includegraphics[width=1\textwidth, angle=0]{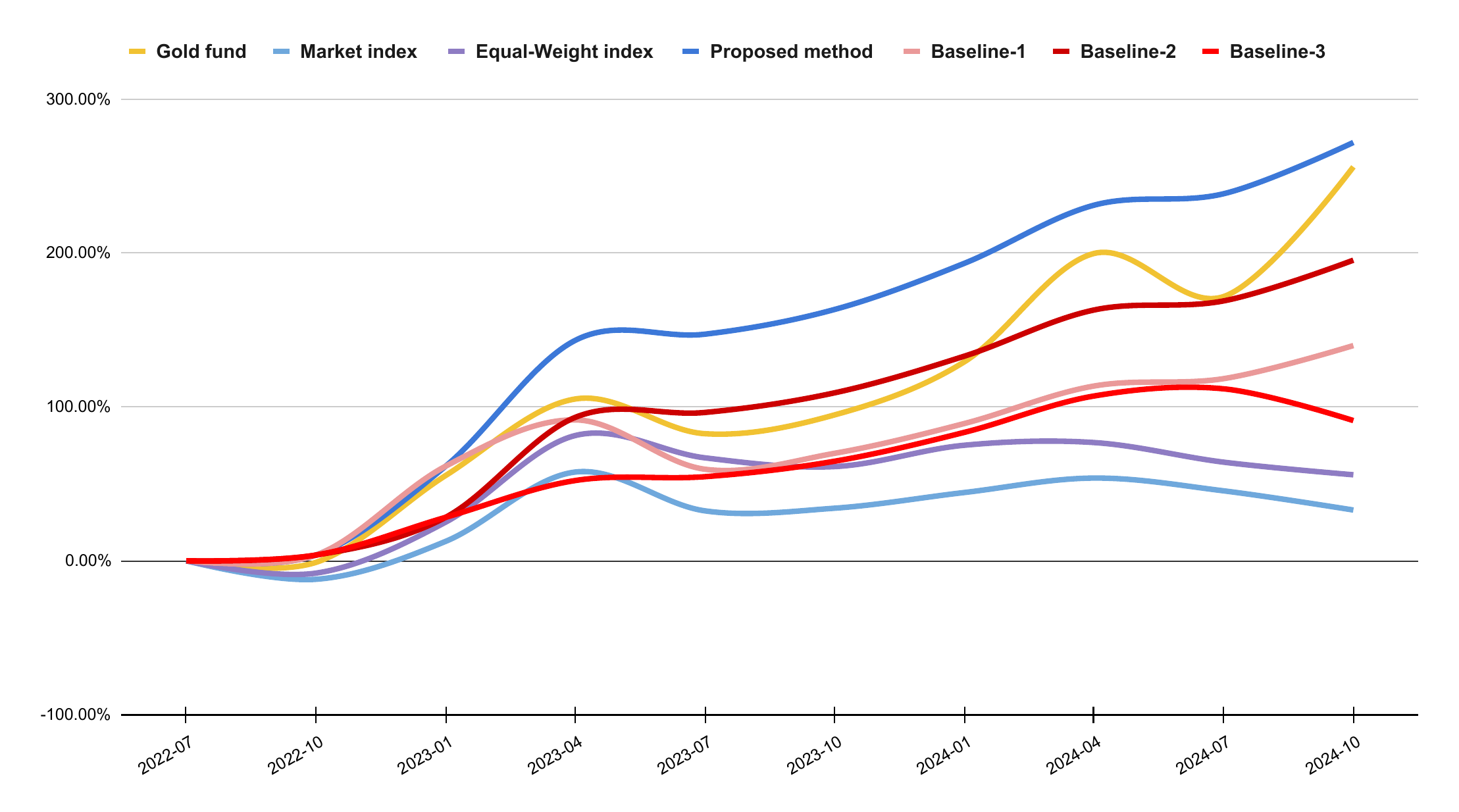}
  \caption{Cumulative nominal return of strategies in test period}
  \label{fig:results-cumulative-return}
\end{center}
\end{figure}

\begin{figure}[h!]
\begin{center}
  \includegraphics[width=1\textwidth, angle=0]{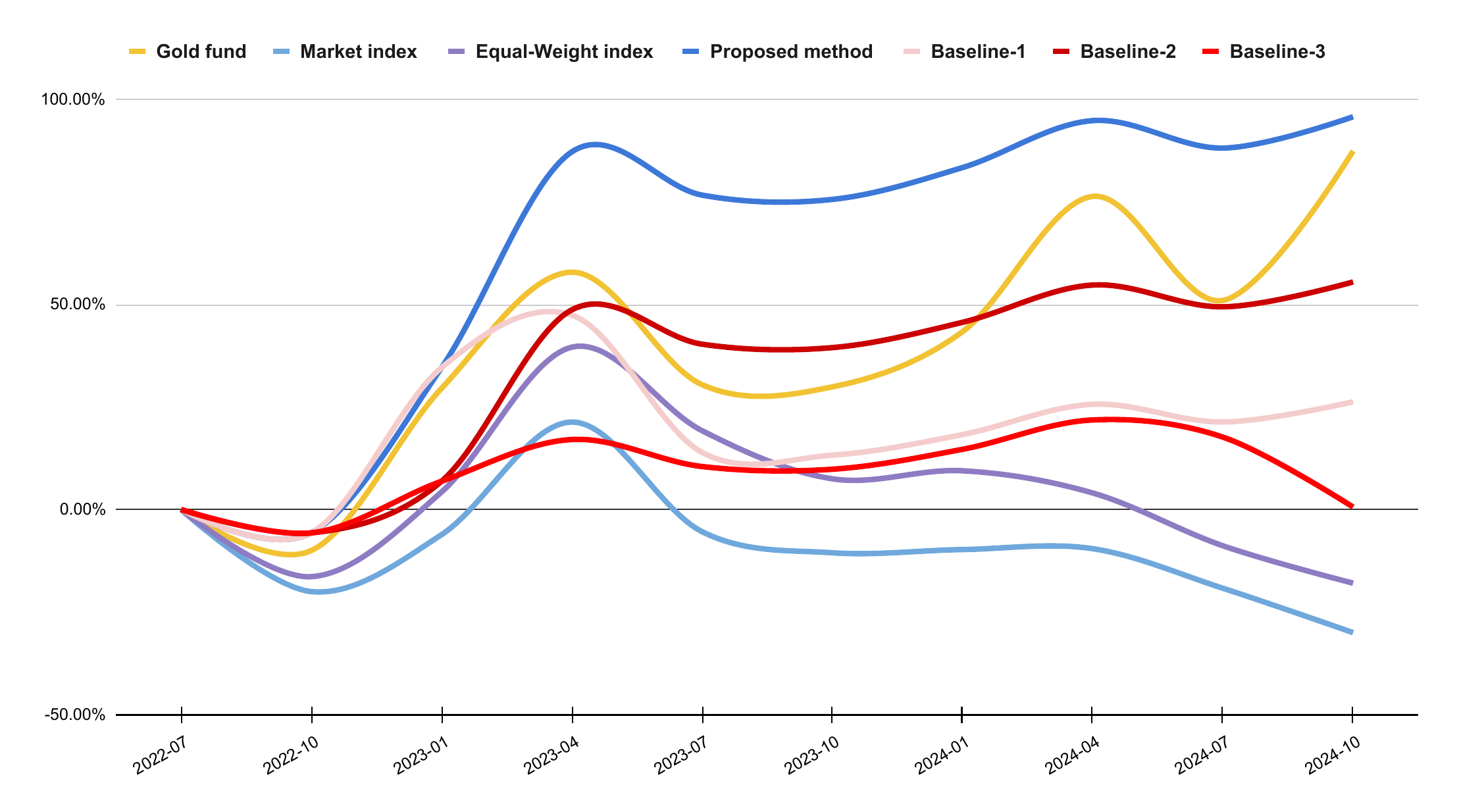}
  \caption{Cumulative real return of strategies in test period}
  \label{fig:results-cumulative-return-adjusted}
\end{center}
\end{figure}

The proposed strategies are based on the fact that \%20 of the portfolio is invested in a stable and low-risk commodity asset such as gold. The remaining fund is invested in the stock market and bond market (as fixed-income ETFs). If the model predicts the stock market return is more than fixed-income, strategy number 1 is followed, and the main part of assets is invested in the stock market. Otherwise, strategy number 2 is followed, and the main part of assets are invested in bond securities.
The three months return of the selected strategies with asset returns are presented in figures \ref{fig:results-3-months-return} and \ref{fig:resutls-3-months-return-adjusted}, which contain nominal and real returns in the specified periods. The cumulative return of the strategies are presented in figures \ref{fig:results-cumulative-return}, and \ref{fig:results-cumulative-return-adjusted}.
As stated, one of the strengths of the proposed approach is that it works well in an inflationary economic condition. Also, we included the real rate of return in the figures, that is, the return adjusted by inflation effect that is presented in \ref{eq:nominal-rate-of-return}.

\begin{equation}\label{eq:nominal-rate-of-return}
\textit{Nominal rate of return} = (1 + \textit{real rate of return}) * (1 + \textit{inflation rate}) - 1
\end{equation}

The global gold price was high during the test period, as seen in Figure \ref{fig:results-cumulative-return-adjusted}, because of the worldwide demand for this commodity. Our proposed method outperforms the main indexes of gold and stock markets, as well as the methods of baselines. This event is because our method can predict stock market direction with high accuracy. Changing assets into lower-risk bonds can prevent capital losses and gain a moderate return. And in the stock market uptrend periods, it can achieve good returns and take advantage of both the commodities and stock market return.

\section{Conclusion} \label{section:conclusion}

This paper introduced an end-to-end decision support system that aids investors in predicting stock returns based on fundamental data analysis. Fundamental data analysis is a widely recognized and used approach that is more powerful than other methods, such as technical analysis. However, it has more challenging tasks in gathering, cleaning, and modeling the fundamental data. We tackled these problems by designing a robust decision support system that pervasively covered the mentioned stages. Also, another main strength of the proposed approach is in the feature engineering stage, which includes various categories of features. The most crucial feature categories are financial ratio, stock type, stock trading features, and also macro-economic conditions of the country. These modeling of data have superior accuracy in high inflation economies compared to novel fundamental analysis approaches. 

For a real-world application of the proposed approach, we predicted stock market direction and designed multiple strategies for asset allocation in various market trend orientations. By including bond and gold assets in the portfolio, the gained return increased by moving capital to lower-risk securities during stock market downtrends. Therefore, the critical success factor was active portfolio management and rebalancing at good times. The proposed method has many applications in different areas, such as finance, investing, government decision-making, and others. Also, individual and institutional investors can benefit from its real-time performance by processing vast amounts of fundamental data.

This work can be used as a groundwork for future researches. The model can consider the investor risk level and propose a customized portfolio concerning the selected level of risk. Also, the model can propose different types of assets, such as real estate, cash, other types of commodities, etc. This model has great achievement in mid- and long-term horizons, but it should be analyzed more in short-term time horizons. Also, other future works include forecasting commodities with a custom model, using other types of novel machine learning models, and evaluating the method on other countries' markets.

In conclusion, our results showed that using a comprehensive decision support system for predicting stock return based on vast informational data, such as stock fundamental data, improves the accuracy of predictions and has many applications in different areas, such as asset allocation and investing for investors.

\bibliographystyle{plain} % We choose the "plain" reference style
\bibliography{refs_introduction,refs_literature_review,refs_others}

\end{document}